\documentclass[11pt, letter]{article}

\usepackage{stolenstyle}

\usepackage{amsmath,epsf,amssymb,latexsym,amsthm,setspace,bbm,array,pifont,enumerate}

\usepackage[matrix,arrow,frame,import,curve,color]{xy}

\usepackage{tikz}

\usetikzlibrary[shapes.geometric]

\tikzset{surcirc/.style={shape=circle,inner sep=3pt,draw}}
\tikzset{surcirc2/.style={shape=circle,inner sep=1.5pt,draw}}

\DeclareFontFamily{U}{rsf}{}
\DeclareFontShape{U}{rsf}{m}{n}{
  <5> <6> rsfs5 <7> <8> <9> rsfs7 <10-> rsfs10}{}
\DeclareMathAlphabet\Scr{U}{rsf}{m}{n}

\def\CO#1#2{{[#1,#2]}}
\def\AC#1#2{{\{#1,#2\}}}

\def\iden{{\mathbbm 1}}

\def\rep#1{{{\boldsymbol{#1}}}}

\def\C{{\mathbb C}}

\def\R{{\mathbb R}}
\def\Z{{\mathbb Z}}

\def\diag{\operatorname{diag}}

\def\SO{\operatorname{SO}}
\def\SL{\operatorname{SL}}

\def\SU{\operatorname{SU}}

\def\Sp{\operatorname{Sp}}
\def\Spin{\operatorname{Spin}}
\def\GG{\operatorname{G}}

\def\GE{\operatorname{E}}

\def\so{\operatorname{\mathfrak{so}}}

\def\su{\operatorname{\mathfrak{su}}}
\def\Lu{\operatorname{\mathfrak{u}}}

\def\p{\partial}
\def\pb{\bar{\partial}}

\def\la{\langle}
\def\ra{\rangle}

\def\ff#1#2{{\textstyle\frac{#1}{#2}}}

\def\cA{{\cal A}}

\def\cC{{\cal C}}

\def\cF{{\cal F}}
\def\cG{{\cal G}}

\def\cJ{{\cal J}}
\def\cK{{\cal K}}

\def\cN{{\cal N}}

\def\cP{{\cal P}}
\def\cQ{{\cal Q}}

\def\cX{{\cal X}}

\def\cZ{{\cal Z}}

\newcommand\vphi{\varphi}

\newcommand\Phit{\widetilde{\Phi}}

\newcommand\cb{\overline{c}}

\newcommand\hb{\overline{h}}

\newcommand\zb{\overline{z}}

\newcommand\bt{\widetilde{b}}

\newcommand\Gt{\widetilde{G}}

\newcommand\Lt{\widetilde{L}}

\newcommand\Tt{\widetilde{T}}

\newcommand\Xt{\widetilde{X}}

\def\bea#1\eea{\begin{align}#1\end{align}}

\def\bes #1\ees{\begin{split}#1\end{split}}

\newcommand{\be}{\begin{equation}}
\newcommand{\ee}{\end{equation}}

\def\bT{{\boldsymbol{T}}}

\def\bg{{\boldsymbol{g}}}

\def\bG{{\boldsymbol{G}}}
\def\bJ{{\boldsymbol{J}}}
\def\bj{{\boldsymbol{j}}}
\def\bg{{\boldsymbol{g}}}
\def\bt{{\boldsymbol{t}}}

\def\bq{{\boldsymbol{q}}}

\def\bpsi{{\boldsymbol{\psi}}}

\def\preprint{  {{EFI-17-15}}\\ {{IPhT-T17/103}}  \\ NSF-ITP-17-082}

\title{Spacetime supersymmetry in low-dimensional perturbative heterotic compactifications}
\author[a] {Ilarion V.~Melnikov,}
\author[b] {Ruben Minasian,}
\author[\,c] {and Savdeep Sethi}
\affiliation[a]{Department of Physics and Astronomy, James Madison University, VA 22807, USA}
\affiliation[b]{  Institut de Physique Th\'eorique, Universit\'e Paris Saclay,  CNRS, CEA\\ 
F-91191 Gif-sur-Yvette, France}
\affiliation[c]{Enrico Fermi Institute, University of Chicago, Chicago, IL 60637, USA}
\emailAdd{melnikix@jmu.edu}
\emailAdd{ruben.minasian@cea.fr}
\emailAdd{sethi@uchicago.edu}
\abstract{We study the constraints of spacetime supersymmetry for perturbative three-- and two--dimensional 
Minkowski vacua of the critical heterotic string.  Assuming a standard RNS construction of the spacetime supersymmetry generators and a compact unitary internal superconformal worldsheet theory, we describe the worldsheet structures associated to various spacetime supersymmetries.  In three dimensions we show that there are no CFT surprises:  each allowed spacetime supersymmetry is realized by a supergravity compactification.  As a recent orbifold construction shows, in two dimensions there are more exotic possibilities,  and we discuss how these fit into our analysis.}

\begin{document}

\maketitle

\section{Introduction} \label{s:intro}
While it remains to be seen whether string theory can predict the weak mixing angle, we know that the internal structure of a string vacuum is constrained by coarse properties of the spacetime, such as its dimension and symmetries.  A classic example is provided by a perturbative critical string compactification, where the worldsheet theory is taken to be a (suitably supersymmetric) sigma model on $\R^{1,D-1}$ tensored with a compact unitary conformal theory and a ghost sector.  In the bosonic string we have a basic requirement on the internal CFT:  its central charge must be $c= 26-D$.  In the superstring, where the worldsheet theory is coupled to (1,1) supergravity, or the heterotic string, where we couple to (1,0) supergravity the critical dimension is $15$, and the internal theory has $c=15-D$.

Spacetime supersymmetry leads to further constraints on the internal theory.  For instance, for a Ramond--Neveu--Schwarz (RNS) heterotic string with a Friedan-Martinec-Shenker (FMS) covariant construction of the fermion vertex operators, spacetime supersymmetry requires the internal worldsheet theory to have enhanced superconformal invariance~\cite{Banks:1987cy}.
 Let us denote the holomorphic worldsheet superconformal algebra (SCA) of a unitary compact CFT with central charge $c$, N supercharges and a unique spin $2$ current by $\cA^{\text{N}}_c$.\footnote{There is a  classification~\cite{Ramond:1976qq,Spindel:1988sr,Sevrin:1988ew} of the SCAs that only have currents with spins in $\{0,1/2,1,3/2,2\}$ and a unique spin 2 current.  All such SCAs have $N\le 4$, and those with $N \le 3$ are unique up to isomorphism; for $N=4$ our notation $\cA^{4}_c$ denotes the so-called ``small'' N=4 algebra.}  The $D=4$ results are then well-known, and we summarize them next.

 The RNS string requires the internal theory to have an $\cA^{1}_9$ SCA.   The vacuum will have $\cN=1$ $D=4$ supersymmetry if and only if $\cA^{1}_9 \subset \cA^{2}_9$, and all NS states carry integer charge with respect to the R-symmetry of $\cA^{2}_9$.   $\cN=2$ spacetime supersymmetry requires $\cA^{2}_9 \subset \cA^{4}_6\oplus \cA^{2}_3$, while $\cN=4$ in spacetime requires $\cA^{2}_9 \subset \cA^{2}_3\oplus \cA^{2}_3\oplus \cA^{2}_3$; each $\cA^2_3$ factor has integral R-charges and corresponds to the holomorphic sector of a sigma model with a $T^2$ targetspace.

Each of the enhancements has a simple geometric realization as a weakly coupled non-linear sigma model that can be studied in a supergravity limit.  For instance, Calabi-Yau compactification with standard embedding leads to a holomorphic $\cA^{2}_9$ SCA.\footnote{The investigations of the worldsheet structure of Calabi-Yau compactifications and related theories~\cite{Hull:1985jv,Sen:1986mg} showed that $\cA^{2}_9$ SCA with integral R-charges was a sufficient condition for spacetime supersymmetry, and the work of~\cite{Banks:1987cy} went on to show this to be necessary as well.}  To enhance supersymmetry, we can specialize the Calabi-Yau manifold to $\text{K3}\times T^2$, which leads to $\cA^{4}_6\oplus \cA^{2}_3$ and $\cN = 2$, or  we may choose it to be $T^6$, which leads to $\cA^{2}_3\oplus\cA^{2}_3\oplus \cA^{2}_3$ and $\cN=4$. 

In these geometric examples of enhanced supersymmetry the superconformal algebra decomposes because the entire CFT decomposes.  This need not be the case:  there are $\cN=2$ vacua where the target space is a principal $T^2$ bundle over $\text{K3}$---see, e.g.~\cite{Dasgupta:1999ss,Goldstein:2002pg,Becker:2006et,Becker:2009df}---and the worldsheet theory does not factor into a product of two CFTs, even though the SCA does decompose as required.  Indeed, by demanding that the requisite worldsheet structure is realized in a non-linear sigma model it is possible to show that these are the most general geometric solutions with $\cN=2$ spacetime supersymmetry~\cite{Melnikov:2012cv}.

In this note we generalize these familiar results to $D =3$ and $D=2$.  We focus on perturbative heterotic vacua where the spacetime CFT decomposes into a free ghost sector with $(c,\cb) = (-15,-26)$, a superconformal $R^{1,D-1}$ SCFT for the Minkowski directions, and a unitary compact ``internal'' CFT.  Similar results were discussed in the context of Narain/lattice compactification in~\cite{Lerche:1988np,Ferrara:1989nm,Ferrara:1989ud}, where spacetime supersymmetry was linked to exceptional Kac-Moody symmetries of the full matter SCFT.  In this work we go beyond those necessary conditions and relate spacetime supersymmetry to algebraic structures of the internal unitary SCFT.  

We find that for $D=3$ every spacetime supersymmetry algebra that can be realized by this construction has a geometric realization, i.e. where the internal theory is a weakly coupled non-linear sigma model.  Similarly for $D=2$ every allowed spacetime supersymmetry algebra $\cN = (p,q)$ with both $p$ and $q$ non-zero has a geometric realization, as do $\cN=(p,0)$ algebras with $p\le 4$.  

On the other hand, the $D=2$ case does have surprises, as illustrated in the orbifold theories constructed in~\cite{Florakis:2017zep}.  In that work, the authors construct two orbifolds of $T^8$ with $\cN= (6,0)$ and $\cN=(16,0)$ spacetime supersymmetry.
In an earlier version of this paper we mistakenly discarded such possibilities.  Fortunately, the explicit examples of~\cite{Florakis:2017zep} showed us the errors of our ways, which we now correct.  The worldsheet structure we uncover for $\cN = (p,0)$ with $p>4$ is consistent with the orbifold constructions.  It would be very interesting to determine if all such vacua arise as orbifolds of $T^8$, but our results, while providing some light on the issue, do not establish it one way or the other.

\subsection*{Geometric realizations and general CFT vacua}
To describe our results, we will now discuss the possible geometric realizations in more detail.  Let us consider heterotic compactification to $D=3$ or $D=2$ dimensions.  Of course every one of the former leads to a special case of the latter by compactification on an extra circle.  Suppose further that the CFT is represented by a weakly coupled non-linear sigma model with targetspace an eight-dimensional manifold $X_8$.  Since we are in the heterotic string, we also need to consider a principal bundle $\cP \to X_8$ to describe the $\GE_8\times\GE_8$ or $\Spin(32)/\Z_2$ gauge fields with curvature $\cF$.   It is well-known that spacetime supersymmetry requires a reduction of structure for $X_8$---see, for instance,~\cite{Gauntlett:2003cy}\ for a comprehensive discussion that includes non-trivial torsion.   A simple class of examples is obtained when $X_8$ is a special or exceptional holonomy manifold with holonomy group $G \subset \SO(8)$ and $\cF$ satisfies a generalization of the Hermitian Yang-Mills equations.  These examples lead to the following pattern of spacetime supersymmetries in $D=2$.
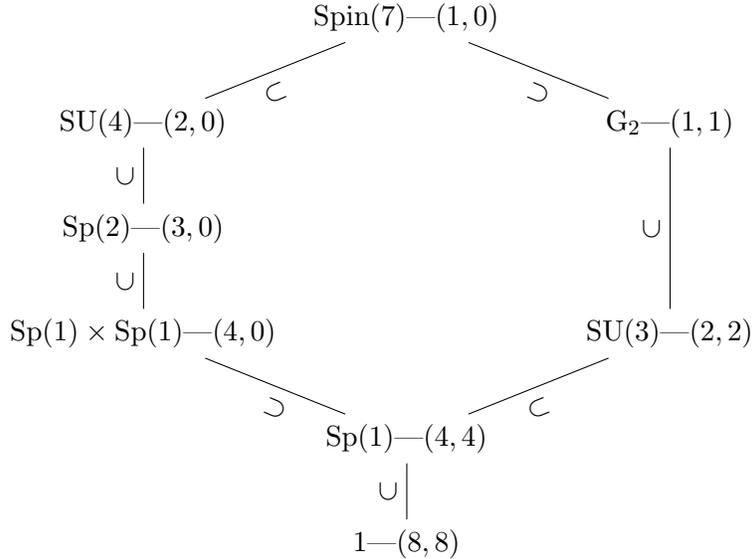
\begin{align*}
\begin{tikzpicture}[scale=0.7]
\draw  (0,0) node (A) {$\Spin(7)$---$(1,0)$};
\draw  (-5,-2) node (B) {$\SU(4)$---$(2,0)$};
\draw (5,-2)  node (C) {$\GG_2$---$(1,1)$};
\draw[thin] (A) --node [anchor=north] {\rotatebox{23}{$\subset$}} (B) ;
\draw[thin] (A) --node [anchor=north] {\rotatebox{157}{$\subset$}} (C) ;
\draw (-5,-4) node (D) {$\Sp(2)$---$(3,0)$};
\draw (5,-6)  node (E) {$\SU(3)$---$(2,2)$};
\draw[thin] (B) --node [anchor=east] {\rotatebox{90}{$\subset$}} (D); 
\draw[thin] (C) --node [anchor=east] {\rotatebox{90}{$\subset$}} (E); 
\draw (-5,-6) node (F) {$\Sp(1)\times\Sp(1)$---$(4,0)$};
\draw[thin] (D) --node [anchor=east] {\rotatebox{90}{$\subset$}} (F); 
\draw (0,-8) node (G) {$\Sp(1)$---$(4,4)$};
\draw[thin] (F)--node [anchor=north] {\rotatebox{157}{$\subset$}} (G);
\draw[thin] (E)--node [anchor=north] {\rotatebox{22}{$\subset$}} (G);
\draw (0,-10) node (H) {$1$---$(8,8)$};
\draw[thin] (H) --node [anchor=east] {\rotatebox{90}{$\subset$}} (G); 
\end{tikzpicture}
\end{align*}
Each entry denotes the holonomy group of $X_8$ and the corresponding \textit{spacetime} supersymmetry.  Each of these geometries yields a non-linear sigma model for the internal CFT with $c=12$ and $\cb = 24$, and because we are working under the auspices of RNS and FMS, the worldsheet theory realizes at least an N=1 superconformal algebra on the holomorphic (supersymmetric) side of the heterotic string---this is denoted by $\cA^{1}_{12}$ in the notation introduced above.  In each of these geometric realizations the worldsheet algebra is enhanced as follows:\begin{align*}
\arraycolsep=7.0pt
\begin{array}{ccccccccc}
G	& \Spin(7)				&\SU(4)		&\Sp(2)		&\GG_2				&\SU(3)	&\Sp(1)^2	&\Sp(1)  & 1 \\[2mm]
\cN 	& (1,0)     				&(2,0) 		&(3,0)		&(1,1)				&(2,2)	& (4,0)			&(4,4)	& (8,8)\\[2mm]
\text{SCA} 	& \cA_{12,\text{I}}^1	&\cA_{12}^2	&\cA_{12}^4		&\cA^1_{12,\text{I}^3}	&\cA^2_{9}\oplus\cA^2_{3}	&\cA^4_6\oplus\cA^4_6 	&\cA^4_6\oplus(\cA^2_3)^{\oplus 2}     & (\cA^2_3)^{\oplus 4}	\\
\end{array}
\end{align*}
Let us make a few comments on this table.
\begin{enumerate}
\item  In the $\Spin(7)$ and $\GG_2$ cases we introduced additional labels.  In these cases, as shown for the weakly coupled worldsheet theory in~\cite{Shatashvili:1994zw}, the theory has an additional structure:  the Virasoro algebra $\cA^0_{12}$ decomposes into a sum of two commuting sub-algebras; in the $\Spin(7)$ case one of the summands is the Ising model; in the $\GG_2$ case one of the summands is the tri-critical Ising model.   

\item Each $N\ge 2$ SCA is extended because the R-charge spectrum is integral.  This means there are holomorphic chiral primary operators that are isomorphic to the identity via spectral flow with R-charge $c/3$ and weight $c/6$.

\item  Each of the $\cA^2_3$ terms arises from a $T^2$ factor in the geometry, and the algebra has a Sugawara decomposition in terms of a free Weyl fermion and two abelian currents. 

\item This table also implies a classification for $D>2$.  We will discuss the $D=3$ case in detail below, but the basic idea is clear:  given any $D>2$ compactification, we can take it down to $D=2$ by compactifying on $T^{D-2}$, and the higher dimensional supersymmetry descends in a natural fashion to one of the non-chiral entries in the table.  For instance with $D=6$ there are just two options---we preserve $8$ supercharges if the worldsheet theory has a $\cA^4_{6}$ SCA, and we obtain $16$ supercharges if the worldsheet theory has a $(\cA^2_3)^{\oplus 2}$ SCA.

\end{enumerate}
We show that perturbative heterotic D=2 compactification with some supersymmetry must belong to one of the following possibilities for the internal CFT.
\begin{enumerate}[i.]
\item  $\cN$ is one of the entries in the second row of the table, in which case the internal CFT has the corresponding SCA.   In each case with $N\ge 2$ the R-charge spectrum is integral, and therefore each $\cA^2_3$ algebra has a Sugawara decomposition into a free chiral primary Weyl fermion and two abelian holomorphic currents.
\item $\cN = (p,0)$ with $p>4$, in which case the worldsheet theory has $\cA^{4}_{6}\oplus \cA^{4}_6$ SCA, and a level $1$ Kac-Moody algebra $\widehat{\so(p)}_1$ that contains the $\widehat{\su(2)}_1\oplus \widehat{\su(2)}_1$ R-symmetry:
\begin{align*}
\widehat{\so(p)}_1 \supset \widehat{\su(2)}_1\oplus \widehat{\su(2)}_1\oplus \widehat{\so(p-4)}_1~,\nonumber\\
\wedge^2\rep{p} = (\rep{3},\rep{1},\rep{1}) \oplus (\rep{1},\rep{3},\rep{1})\oplus (\rep{2},\rep{2},\rep{p-4})~.
\end{align*}
The currents in the $ (\rep{2},\rep{2},\rep{p-4})$ representation lie in short multiplets with respect to each of the $\cA^4_6$ factors.  The central charge $c=12$ and unitarity lead to a bound $p\le 24$.

\end{enumerate}
To obtain the $D=3$ results, we impose an additional requirement on the internal $N=1$ SCFT:  it should decompose into a sum of two terms, one of which is a free $c=3/2,\cb=1$ theory representing the $S^1$.  Moreover, we know that any such theory must have $\cN = (p,p)$ supersymmetry in $D=2$.  Taking a look at those entries in the $D=2$ table and factoring out the circle,
we find the analogue of the previous table for $D=3$ supersymmetry:
\begin{align*}
\arraycolsep=7.0pt
\begin{array}{cccccc}	
G		&\GG_2					&\SU(3)						&\Sp(1)  & 1 \\[2mm]
\cN 		&1						&2							   &4		& 8\\[2mm]
\text{SCA}	&\cA^1_{21/2,\text{I}^3}&\cA^2_{9}\oplus\cA^1_{3/2,S^1}	&\cA^4_6\oplus(\cA^1_{3/2,S^1})^{\oplus 3}    & (\cA^1_{3/2,S^1})^{\oplus 7}\\
\end{array}
\end{align*}
The first line denotes targetspace holonomy groups for geometric realizations; the second line lists all spacetime supersymmetries  (our $\cN$ counts Majorana supercharges in $D=3$) that can be realized in the perturbative heterotic string, and the last line lists the requisite worldsheet structure.  Finally, the $\cA^1_{3/2,S^1}$ denotes the $N=1$ superconformal algebra of a single free Majorana fermion and its superpartner current.  These constraints are consistent with heterotic supergravity results obtained in~\cite{Melnikov:2017wcf}.  

As in compactifications to $D=4$, the results for $D=2$ and $D=3$ just reflect the structure of the holomorphic side of the worldsheet---in particular, a decomposition of the superconformal algebra does not require the whole CFT to decompose.  In geometric terms the generic situation corresponds to a principal torus fibration; when the fibration is trivial the worldsheet theory decomposes. The geometric approach allows an alternative and illuminating way of viewing these constructions:  we first reduce the theory on some torus $T^k$ and then compactify the resulting theory further on a base manifold $X$.  This point of view can be used to identify some features that are qualitatively different for different choices of $D$.    For instance, a $T^2$ fibration over a K3 manifold can lead to either $\cN=2$ or $\cN = 1$ in $D=4$~\cite{Dasgupta:1999ss,Goldstein:2002pg,Becker:2006et}.  On the other hand, we cannot construct a principal $S^1$ fibration over a $\GG_2$ manifold to preserve $\cN=(1,0)$ in $D=2$---this is not obvious in the CFT language, but geometrically it is clear:  the $9$--dimensional theory is non-chiral, so that a further compactification cannot lead to $\cN=(1,0)$.

\subsubsection*{Comments on the result}
This application of 1980s technology has bearing on topics of more modern interest.  First, it turns out that for $D=3$ geometry is a perfect guide for spacetime SUSY in more general CFT vacua.  There may be CFT vacua without any geometric realization, but if this is the case, supersymmetry is not a sufficient criterion to distinguish such interesting solutions.  Second, this leads to additional insight into the structure of heterotic supergravity:  for instance, given a torsional geometry $X_7$ that preserves $\cN \ge 4$ supersymmetry in $D=3$, it is not possible to choose a background gauge field to break supersymmetry further to $\cN = 3$, but the argument is a bit intricate~\cite{Melnikov:2017wcf}.  We now understand why this is the case independently of any details of the geometry, and we can use our result to identify corners in the type II/M/F-theory landscape that cannot have standard perturbative heterotic duals.  One broad class of such theories is M-theory compactified on $X_8$ with $\Sp(2)$ holonomy, which naturally leads to $\cN = 3$ in $D=3$.  In fact, it is possible to find $\cN=3$ $D=3$ M-theory vacua with $X_8 = \text{K3}\times\text{K3}$ and an appropriate choice of $G$-flux~\cite{Melnikov:2017wcf}.

In $D=2$ compactifications there is a class of examples that realize spacetime supersymmetries that cannot be obtained from weakly coupled non-linear sigma models: these are the theories with $\cN = (p,0)$ and $p>4$.  Classifying these is an interesting open problem:  are all such theories orbifolds of $T^8$?  is there a CFT for every value $p\le 24$?  Our work identifies the basic structure that every such theory must possess without making reference to any explicit construction method.

Even in $D=4$ compactifications perturbative type II string theory provides examples of more exotic supergravities, e.g. with $\cN = 3$ or $\cN=5$.  This may appear surprising at first sight, since we may apply the RNS/FMS construction to the superstring as well.  The loophole is that, unlike in the heterotic worldsheet, in type II string spacetime supersymmetries arise from both the holomorphic and anti-holomorphic sectors, and the mixing of these provides for more exotic possibilities.  The supergravity constructions of such vacua involve Ramond-Ramond fluxes and are best understood as vacua of various gauged supergravities.\footnote{A nice recent discussion may be found in~\cite{Trigiante:2016mnt}.}  Their worldsheet description requires a non-RNS approach, such as the Berkovits pure spinor  formalism~\cite{Bedoya:2009np}.  One can also construct exact CFT descriptions as asymmetric orbifolds of standard (2,2) worldsheet theories; see, e.g.~\cite{Ferrara:1989nm,Blumenhagen:2016axv,Blumenhagen:2016rof}.

All such type II vacua with ``exotic'' (from the heterotic CFT realization point of view) supersymmetry cannot have a weakly coupled heterotic dual, despite the fact that many of their close cousins do have relatively simple dual descriptions.  For instance, the $\cN=3$, $D=3$ M-theory vacua on $X_8 = \text{K3}\times\text{K3}$ are closely related to heterotic vacua with well-known perturbative duals.  Of course ``close'' here is misleading --- the difference is measured by a choice of quantized $G$-flux!  Still, perhaps considerations along these lines will teach us something new about the classic subject of string duality.

We should also mention the possibility that some alternative, say Green-Schwarz-like or Berkovits construction of the heterotic worldsheet may lead to additional choices for spacetime supersymmetry.  It would be interesting to find such descriptions, since they may also apply to other heterotic no go theorems such as~\cite{Kutasov:2015eba}.  From the supergravity no-go results of~\cite{Melnikov:2017wcf} we know that to apply to the puzzles raised here, these constructions would have to be inherently stringy.

Finally, we recall that the term ``$D=2$ compactification'' deserves its quotes:  the $D=2$ IR divergences mean that it is impossible to fix the expectation values of scalars that correspond to the CFT moduli---they must be integrated over, and the dynamics of the resulting non-linear sigma model is quite complicated.   Thus, such theories are not easy to understand at finite string coupling. 

Having explained our results and discussed some of their significance, we now turn to the proof.  In section~\ref{s:FMSRNS} we review the set-up of the perturbative heterotic RNS worldsheet and state our assumptions; with these in hand we present the worldsheet realization of the spacetime supersymmetry algebra.  Next, in section~\ref{s:pzero} we consider constraints on realization of $\cN = (p,0)$ spacetime supersymmetry in $D=2$, and in section~\ref{s:pq} we turn to the the case of $\cN = (p,q)$ supersymmetry with $p \ge 2$ and $1 \le q \le p$.  It turns out that $\cN=(1,1)$ and the appearance of the tri-critical Ising model require a slightly different line of development, and we handle that in the last section.

\acknowledgments IVM would like to thank the KITP Scholars program for providing a wonderful environment in which to complete this work.  We thank W.~ Lerche, D.~Morrison, R.~Plesser, D.~Robbins, and S.~Theisen for useful discussions.  We are supported, in part, by 4-VA Initiative ``Frontiers in string geometry,'' and JMU Faculty Assistance Grant, as well as by the National Science Foundation under Grant No. NSF PHY11-25915 (IVM), Agence Nationale de la Recherche under grant 12-BS05-003-01 (RM), and NSF Grant No. PHY-1316960 (SS).

\section{Spacetime supersymmetry from the RNS worldsheet} \label{s:FMSRNS}
Following FMS~\cite{Friedan:1985ge}, we take the following as necessary requirements for a perturbative $\R^{1,1}$ vacuum of the critical heterotic RNS string.  

The worldsheet theory in conformal gauge is a conformal theory with $N=1$ superconformal invariance on the left--moving side of the string.  The degrees of freedom consist of three decoupled theories.
\begin{enumerate}
\item The ghost sector has $(c,\cb) = (-15,-26)$ and is represented by the standard $bc$--$\beta\gamma$ system.  
\item The $\R^{1,1}$ sector, which consists of two bosons $X^\mu$ and their superpartners---a pair of left-moving Majorana-Weyl fermions  $\psi^\mu$.
\item The internal sector, which is a  $(c,\cb) = (12,24)$ unitary and compact $N=1$ superconformal theory.\footnote{Compactness of the CFT means that the theory has a normalizable $\SL(2,\C)$ vacuum and a finite number of primary operators with scaling dimension below any fixed value.}
\end{enumerate}
We work on a Euclidean worldsheet with coordinates $z$ and $\zb$, with the former corresponding to the left-moving side of the string.  We now list some standard OPEs for the free fields.  Throughout we will use a shorthand where we denote the position of a local operator by a subscript; for example $X^\mu_1 = X^\mu(z_1,\zb_1)$.  We also bosonize the $\beta\gamma$ system via $:\beta\gamma: \sim \p \vphi$.  Whenever it is not likely to cause confusion, we will drop the normal ordering symbols for composite operators constructed from free fields.   With that notation we have the familiar
\begin{align}
X^\mu_1 X^\nu_2 &\sim -\ff{\alpha'}{2} \eta^{\mu\nu} \log |z_{12}|^2~, &
\psi^\mu_1 \psi^\nu_2 & \sim \frac{\eta^{\mu\nu}}{z_{12}}~, &
(e^{s\vphi})_1 (e^{t\vphi})_2 & = z_{12}^{-st} : (e^{s\vphi})_1 (e^{t\vphi})_2 :~.
\end{align}
Recall that $e^{s\vphi}$ has weight $h = -s -s^2/2$.

We further assume that our theory has a fermion number operator $e^{i\pi F}$ that can be used to implement a GSO projection to $e^{i\pi F} = 1$.  The internal theory is taken to have an NS sector with some standard properties:  it contains the identity; the NS--NS OPE is single-valued and closes on NS operators, and there is a spin--statistics relation:  the spin of an operator with weights $h,\hb$ satisfies $h-\hb \in \ff{1}{2} \Z$ and
\begin{align*}
h-\hb \in \begin{cases}   \ff{1}{2} + \Z & \text{for fermionic operators} \\
				     \Z			& \text{for bosonic operators}~.
\end{cases}
\end{align*}
Finally, we take the action of $e^{i\pi F}$ on the ghosts and $\R^{1,1}$ fermions to be
\begin{align}
e^{i\pi F} e^{-s\vphi} e^{-i\pi F} & = e^{-i\pi s} e^{-s\vphi}~,&
e^{i\pi F} \psi^\mu e^{-i\pi F} & = - \psi^\mu~.
\end{align}
These are just necessary conditions for the existence of a critical string vacuum---there are additional conditions that involve GSO projections in the right-moving (non--supersymmetric) sector and constraints of modular invariance.  However, this structure will be sufficient for our purposes.

\subsection{The Ramond sector and spacetime supercharges} \label{ss:ramond}
Continuing to follow FMS, we now introduce the Ramond sector and corresponding spin fields.  In the $\R^{1,1}$ sector this is a simple story.  We choose a basis $\{\Gamma^0,\Gamma^1\}$ for the $D=2$ Dirac matrices as follows:
\begin{align}
\Gamma^0 &= \begin{pmatrix}
0 & 1 \\ -1 & 0 
\end{pmatrix}~,&
\Gamma^1 &= \begin{pmatrix}
0 & 1 \\ 1 & 0 
\end{pmatrix}~.
\end{align}
This satisfies $\AC{\Gamma^\mu}{\Gamma^\nu} = 2\eta^{\mu\nu}$ with $\eta = \diag(-1,1)$.  From these we construct the chirality matrix 
\begin{align}
\Gamma_2 &= \Gamma^0\Gamma^1 = \begin{pmatrix}
1 & 0 \\ 0 & -1 
\end{pmatrix}~.
\end{align}
The charge conjugation matrix $\cC =\Gamma^0$ satisfies 
\begin{align}
 \cC^2 &= -\iden~,&  \cC \Gamma^\mu \cC^{-1} &= -\left(\Gamma^{\mu }\right)^{T}~,&
\cC^T &= -\cC~,&
(\cC\Gamma^\mu)^T & = \cC\Gamma^\mu~.
\end{align}
The Ramond ground states, and therefore the corresponding spin fields, must furnish a representation of the zero modes $\psi^\mu_0$ and the fermion number.  The smallest such representation is two-dimensional and provided by the Dirac matrices.  The two spin fields $S_a$, with $a=\pm$ the chirality label, have $h=1/8$ and satisfy
\begin{align}
\label{eq:psiS}
\psi^\mu(z) S_a(0) \sim \frac{1}{\sqrt{2z}} S_b(0) \Gamma^\mu_{ba}~,
\end{align}
where the $\sqrt{2}$ is a convenient normalization factor; moreover, conformal invariance and spacetime Lorentz symmetry fix the OPE
\begin{align}
S_a(z_1) S_b(z_2) \sim z_{12}^{-1/4}\left( \cC_{ba} + \sqrt{z_{12}} \ff{1}{\sqrt{2}} (\cC\Gamma_\nu)_{ba} \psi^\nu(z_2) + O(z_{12}) \right)~.
\end{align}
The fermion number acts by
\begin{align}
e^{i\pi F} S_a e^{-i\pi F} = i S_b (\Gamma_2)_{ba}~,
\end{align}
and this is consistent with the OPE structure.

With this framework in place, we construct the massless gravitino vertex operators in the $-1/2$ picture:
\begin{align}
e^{ik\cdot X}\pb X^\mu e^{-\vphi/2} S_+ \Sigma^A_+~,&&
e^{i k \cdot X}\pb X^\mu e^{-\vphi/2} S_- \Sigma^{\dot A}_-~,
\end{align}
where $\Sigma^A_+$ and $\Sigma^{\dot A}_-$ are spin fields of the internal theory and satisfy the following requirements.
\begin{enumerate}
\item They are $N=1$ primary operators with weight $h=1/2$ and $\hb = 0$.  In fact, their OPE with the supercharge of the internal theory must have square root branch cuts and at most a $1/\sqrt{z}$ singularity~\cite{Banks:1987cy,Banks:1988yz}.
\item To survive the GSO projection the $\Sigma^A_+$ carry fermion number $+1$, while the $\Sigma^A_-$ carry fermion number $-1$.
\end{enumerate}
These operators are trivial on-shell---the two-dimensional gravitino does not propagate.  However, by setting $k=0$ and stripping off the right-moving $\pb X^\mu$, we obtain candidates for the spacetime supercharges in the $-1/2$ picture:
\begin{align}
\cQ^A_+ &= e^{-\vphi/2} S_+ \Sigma^A_+~,&
\cQ^{\dot A}_- & = e^{-\vphi/2} S_- \Sigma^{\dot A}_-~.
\end{align}
These are holomorphic currents of weight $h=1$, $\hb = 0$, and there are corresponding conserved charges
\begin{align}
Q^A_+ &= \oint \frac{dz}{2\pi i} \cQ^A_+(z)~,&
Q^{\dot A}_- &= \oint \frac{dz}{2\pi i} \cQ^{\dot A}_-(z)~.
\end{align}
In order for these to give rise to spacetime supersymmetries, the OPE of the $\cQ$ should close onto conserved worldsheet currents associated to translations and central charges in the $-1$ picture.  These correspond to, respectively, the currents
\begin{align}
\cP^\mu &= e^{-\vphi} \psi^\mu~, &
\cZ^{A\dot B} & = i e^{-\vphi} \Psi^{A\dot B}~,
\end{align}
where $\Psi^{A\dot B}$ is some NS sector $h=1/2$ holomorphic field in the internal theory.   Integrating these we obtain the charges $P^\mu$ and $Z^{A\dot B}$.    Now, finally, since we know the OPEs of the ghosts and the $\R^{1,1}$ spin fields $S_a$, we can draw conclusions about the OPEs of the $\Sigma$s.    We find that to obtain the $\cN = (p,q)$ supersymmetry algebra
\begin{align}
\AC{Q^A_+}{Q^B_+} &= \frac{\delta^{AB}}{\sqrt{2}} (P^0 + P^1)~,&
\AC{Q^{\dot A}_-}{Q^{\dot B}_-} &= \frac{\delta^{\dot{A}\dot{B}}}{\sqrt{2}} (P^0 - P^1)~,&
\AC{Q^A_+}{Q^{\dot B}_-} & = Z^{A\dot{B}}~,
\end{align}
with $1 \le A\le p$ and $1 \le \dot{A} \le q$
requires
\begin{align}
\label{eq:SUSYconstraint}
\cQ^A_+(z_1) \cQ^B_+(z_2) &\sim \frac{\delta^{AB} (\cP^{0}(z_2) + \cP^{1}(z_2))}{\sqrt{2}z_{12}}~, &
\cQ^{\dot A}_-(z_1) \cQ^{\dot B}_-(z_2) &\sim \frac{\delta^{\dot{A}\dot{B}} (\cP^{0}(z_2) - \cP^{1}(z_2))}{\sqrt{2}z_{12}}~,  \nonumber\\
\cQ^A_+(z_1) \cQ^{\dot B}_- (z_2) & \sim \frac{\cZ^{A\dot B}(z_2)}{\sqrt{2} z_{12}}~,
\end{align}
and this means the $\Sigma$s satisfy
\begin{align}
\label{eq:SigmafromSUSY}
\Sigma^A_+(z_1) \Sigma^B_+(z_2) &\sim \frac{\delta^{AB}}{z_{12}}~,&
\Sigma^{\dot A}_-(z_1) \Sigma^{\dot B}_-(z_2) &\sim \frac{\delta^{\dot A\dot B}}{z_{12}}~,&
\Sigma^A_+(z_1) \Sigma^{\dot B}_-(z_2) & \sim \frac{1}{\sqrt{z_{12}}} \Psi^{A\dot B}_2~.
\end{align}

Before we leave this general discussion, we point out one general fact:  if both $p$ and $q$ are non-zero, then
the $\Psi^{A\dot B}$ cannot all vanish.  To see this, we compute the four-point function
\begin{align}
\la \cQ^A_+(z_1) \cQ^B_+(z_2) \cQ^{\dot A}_-(z_3) \cQ^{\dot B}_-(z_4)\ra~.
\end{align}
This function is completely determined by its singularities, which we know from~(\ref{eq:SUSYconstraint}).   We find
\begin{align}
\la \cQ^A_+(z_1) \cQ^B_+(z_2) \cQ^{\dot A}_-(z_3) \cQ^{\dot B}_-(z_4)\ra~ &=
\frac{\delta^{AB}\delta^{\dot A \dot B}}{z_{12} z_{23} z_{24} z_{34}}
-\frac{\Pi^{B\dot B,A\dot A}}{2z_{13}z_{23} z_{24} z_{34}} 
-\frac{\Pi^{B\dot A,A\dot B}}{2z_{14} z_{23}z_{24} z_{34}}~,
\end{align}
where $\Pi^{A \dot A,B\dot B}$ is defined by the two-point function 
\begin{align}
\Pi^{A\dot A,B\dot B} = z_{12} \la \Psi^{A\dot A}_1 \Psi^{B\dot B}_2\ra~.
\end{align}
However, special conformal invariance of the four-point function requires
\begin{align}
 \Pi^{B\dot B,A\dot A}+\Pi^{B\dot A,A\dot B} = 2\delta^{AB}\delta^{\dot A \dot B}~,
\end{align}
which means that $\Pi^{A\dot A,B\dot B}$, and consequently $\Psi^{A\dot A}$ cannot vanish
when $p$ and $q$ are both non-zero.

\subsection{Two geometric examples} \label{ss:geometryexamples}
We illustrate the preceding discussion with two examples.  First we take the compactification manifold $X_8$ to be a Calabi-Yau four-fold.  The worldsheet theory then has an $N=2$ superconformal invariance and integral R-charges, with the R-current $J^R$ normalized to
\begin{align}
J^R_1 J^R_2 \sim \frac{4}{z_{12}^2}~.
\end{align}
We bosonize it as $J^R = 2 i \p H$, where $H$ is a holomorphic boson with $H_1 H_2 \sim -\log z_{12}$.  The fermion number is identified with $e^{i\pi J^R_0}$, where $J^R_0$ is the R-charge.

There are two spin fields $\Sigma^A_+$, which we combine into a complex field $\Sigma$ and its conjugate $\Sigma^\dag$.  In bosonized form they are given by
\begin{align}
\Sigma & = e^{i H}~, &
\Sigma^\dag & = e^{- i H}~.
\end{align}
These have $h=1/2$,  R-charges $\pm 2$, fermion number $+1$, and the only non-trivial OPE is $\Sigma_1 \Sigma^\dag_2 \sim z_{12}^{-1}$.   As a result, we obtain $\cN = (2,0)$ spacetime supersymmetry.

For our next example, we consider $X_8 = X_6 \times T^2$, where $X_6$ is a Calabi-Yau three-fold.  The $c=12$ $N=2$ SCA is of the form
\begin{align}
\cA^{2}_{12} \subset \cA^{2}_{9}\oplus \cA^{2}_{3}~,
\end{align}
with R-currents decomposing analogously as
\begin{align}
J^R & = \bJ + \bj~, & 
\bJ_1\bJ_2 & \sim \frac{3}{z_{12}^2}~,&
\bj_1 \bj_2 & \sim \frac{1}{z_{12}^2}~.
\end{align}
We bosonize the R-currents as $\bJ = i\sqrt{3} \p H$ and $\bj = i\p h$.  There are now four spin fields:
\begin{align}
\Sigma_+ & = e^{\ff{i}{2} (\sqrt{3}H+h)}~,&
\Sigma_+^\dag & = e^{-\ff{i}{2} (\sqrt{3}H+h)}~,&
\Sigma_- & = e^{\ff{i}{2} (\sqrt{3}H-h)}~,&
\Sigma_-^\dag & = e^{-\ff{i}{2} (\sqrt{3}H-h)}~.&
\end{align}
These have the OPEs
\begin{align}
\Sigma_{\pm 1} \Sigma_{\pm2} &\sim z_{12} (e^{i\sqrt{3}H\pm h})_2~,&
\Sigma_{\pm 1} \Sigma^\dag_{\pm2} &\sim \frac{1}{z_{12}} + \ff{1}{2}(\bJ_2 \pm \bj_2) +O(z_{12}), \nonumber\\
\Sigma_{\pm 1} \Sigma_{\mp2} & \sim \sqrt{z_{12}} (e^{i\sqrt{3}H})_2~,&
\Sigma_{\pm 1} \Sigma^\dag_{\mp 2} & \sim \frac{1}{\sqrt{z_{12}}} (e^{\pm i h})_2 +\ff{1}{2}  \sqrt{z_{12}}  :\!\!(\bJ\pm\bj) e^{\pm i h}\!\!:_2~.
\end{align}
The last one is particularly interesting, since it involves the NS fields $e^{\pm i h}$.  These have $h=1/2$ and $q=\pm 1$; indeed, they are just the bosonized version of the free Weyl fermion of the $T^2$ theory and its conjugate:  they are an instance of the $\Psi^{A\dot{B}}$ fermions in the general discussion above. These spin fields and OPEs lead to $\cN = (2,2)$ spacetime supersymmetry.

From our general discussion it is clear that we did not need a geometric realization of these theories.  In the first case it is sufficient to have a $\cA^{2}_{12}$ SCA with integral R-charges, and in the second case  a $\cA^{2}_9\oplus\cA^{2}_3$ SCA with integral R-charges in both factors.  As soon as this holds, we have spectral flow in each case, and therefore can construct all of the required spin fields as illustrated.  This is a familiar feature of strings based on $N=2$ theories~\cite{Lerche:1989uy,Vafa:1989xc}.

\section{(p,0) spacetime supersymmetry in two dimensions} \label{s:pzero}
We now turn to various possibilities for $\cN = (p,0)$.  For each $p$ our aim is to identify the structure in the holomorphic sector of the theory that is necessary and sufficient to obtain $\cN=(p,0)$ supersymmetry.

Regardless of spacetime supersymmetry, the internal theory has $N=1$ superconformal invariance with a $\cA^{1}_{12}$ SCA.  This is generated by the spin $3/2$ supercharge $G(z)$ and the energy-momentum tensor $T(z)$.  The two satisfy
\begin{align}
G_1 G_2 &\sim \frac{2c/3}{z_{12}^3} + \frac{2 T_2}{z_{12}}~, &
T_1 G_2 &\sim \frac{3/2 G_2}{z_{12}^2} + \frac{\p G_2}{z_{12}}~,&
T_1 T_2 &\sim \frac{c/2}{z_{12}^4} + \frac{2T_2}{z_{12}^2} + \frac{\p T_2}{z_{12}}~.
\end{align}
We also have the corresponding mode expansions and commutation relations:
\begin{align}
\label{eq:N1modes}
\AC{G_r}{G_s}& = \ff{c}{3}(r^2-\ff{1}{4})\delta_{r,-s} + 2 L_{r+s}~,&
\CO{L_m}{G_r} &= (\ff{m}{2}-s) G_{m+s}~,& \nonumber\\
\CO{L_m}{L_n} & = \ff{c}{12}(m^3-m)\delta_{m,-n} + (m-n) L_{m+n}~.
\end{align}
In the NS sector $r \in \ff{1}{2} +\Z$, and $m \in \Z$. 

\subsection{Minimal supersymmetry} \label{ss:minsusy}
In the case of  minimal  $\cN = (1,0)$ spacetime supersymmetry we need a single $h=1/2$ holomorphic spin field $\Sigma$ with OPE $\Sigma_1 \Sigma_2 \sim \ff{1}{z_{12}}$.  This is a familiar object --- a holomorphic Majorana-Weyl fermion.  As is standard, we construct a corresponding energy-momentum tensor $T^\Sigma$
\begin{align}
T^\Sigma = -\frac{1}{2} : \Sigma \p \Sigma :~.
\end{align}
It is then easy to check that $T^\Sigma$ commutes with $(T-T^\Sigma)$, and we conclude that the $c=12$ holomorphic Virasoro algebra decomposes into a sum of a $c=1/2$ algebra and a $c= 23/2$ algebra.  The former is precisely the holomorphic sector of the Ising model.   This is the structure found in the context of $\Spin(7)$ non-linear sigma models~\cite{Shatashvili:1994zw}.  

It is important to remember that $\Sigma$ is a spin field.  It is not in the NS sector, and, in fact, NS sector operators will include the $h=1/16$ spin field $\sigma$ of the Ising model.  Similarly, the ``usual'' fermion number of the Ising model, which assigns fermion number $1$ to $\Sigma$, has nothing to do with the $e^{i\pi F}$ of the internal CFT; as explained in~\cite{Shatashvili:1994zw}, the latter is realized by the Ising $\Z_2$ symmetry $\sigma \to -\sigma$.

\subsection{$\cN = (2,0)$}
Consider next the case of two $\Sigma^A_+$ fields, and define $\Sigma = \ff{1}{\sqrt{2}}(\Sigma^1_+ + i \Sigma^2_+)$, as well as $\Sigma^\dag =\ff{1}{\sqrt{2}}( \Sigma^1_+-i\Sigma^2_+)$.  These are components of a holomorphic Weyl fermion with OPEs
\begin{align}
\Sigma(z) \Sigma^\dag(w) &\sim \frac{1}{z-w}~, &
\Sigma(z) \Sigma(w) & \sim O(z-w)~,&
\Sigma^\dag(z) \Sigma^\dag(w) & \sim O(z-w)~.
\end{align}
We also obtain a current $J_\Sigma = :\Sigma \Sigma^\dag:$ in the NS sector, which assigns charge $+1$ to $\Sigma$ and $-1$ to $\Sigma^\dag$.
This current can be bosonized as $J_\Sigma= i \p H$, and the $\Sigma$, $\Sigma^\dag$ are represented by pure exponentials $e^{\pm i H}$.
Following~\cite{Banks:1987cy}, we observe that we can also bosonize the supercharge
\begin{align}
G = \frac{1}{\sqrt{2}}\sum_{q} G^{(q)} e^{i q H}~.
\end{align}
But, we know that in order for the $\cQ^A_+$ to be physical $G$ has to have a square root branch cut with both $\Sigma$ and $\Sigma^\dag$ and at most $z^{-1/2}$ singularities; that means the only allowed values of $q$ in the sum are $q=\pm 1/2$.  It also follows that $G$ cannot contain $J^R$ or its derivatives, since those would lead to higher poles in the $\Sigma$--$G$ OPE.

Let $J^R = 2 J_\Sigma$, so that the components of $G$ carry charge $\pm 1$ with respect to $J^R$.  Then, 
\begin{align}
J^R(z) J^R(w) \sim \frac{12/3} {(z-w)^2}~,
\end{align}
and $\Sigma = e^{iH}$ and $\Sigma^\dag = e^{-iH}$, just as for the $\text{CY}_4$ discussion above.  This does not demonstrate that we have an $N=2$ SCA, but, following~\cite{Banks:1987cy}, it is easy to show this is the case.

We write 
\begin{align}
G &= \ff{1}{\sqrt{2}} (G^+ + G^-)~, &
G' &= \ff{1}{\sqrt{2}} (G^+ - G^-)~,
\end{align}
so that we have the OPEs
\begin{align}
G_1 J^R_2 &\sim \frac{-G'_2}{z_{12}}~,&
J^R_1 G'_2 & \sim \frac{G_2}{z_{12}}~.
\end{align}
The first of these means $J^R$ is an $N=1$ superconformal primary field.  Any $N=1$ superconformal primary operator $X$ of weight $h$ fits into a multiplet with its superpartner $\cX$ of weight $h+1/2$ with OPEs
\begin{align}
G_1 X_2&\sim \frac{\cX_2}{z_{12}}~,&
G_1 \cX_2 & \sim \frac{2h X_2}{z_{12}^2} + \frac{\p X_2}{z_{12}}~.
\end{align}
Applying this to $J^R$ and $-G'$ with $h=1$, we find
\begin{align}
G_1 J_2& \sim  \frac{-G'_2}{z_{12}}~,&
G_1 G'_2 & \sim -\frac{2J^R_2}{z_{12}^2} -  \frac{\p J^R_2}{z_{12}}~.
\end{align}
Combining all of these facts, we have the following relations
\begin{align}
J^R(z) J^R(w) &\sim \frac{4}{(z-w)^2} &\implies&& \CO{J^R_m}{J^R_n} &=4m\delta_{m,-n}~, \nonumber\\
J^R(z) G(w) &\sim \frac{G'(w)}{z-w} &\implies&& \CO{J^R_0}{G_s} &= G'_s~,\nonumber\\
J^R(z) G'(w) &\sim \frac{G(w)}{z-w} &\implies&& \CO{J^R_0}{G'_s} &= G_s~,\nonumber\\
G(z) G'(w) & \sim -\frac{2J^R(w)}{(z-w)^2} -  \frac{\p J^R(w)}{z-w}&\implies&& \AC{G_r}{G'_s} &= -(r-s) J^R_{r+s}~.
\end{align}
Now we use these and the Jacobi identity to prove $\AC{G'_r}{G'_s} + \AC{G_r}{G_s} = 0$:
\begin{align}
\AC{G'_r}{G'_s} = \AC{G'_r}{\CO{J^R_0}{G_s}} & = \AC{G_s}{\CO{G'_r}{J^R_0}}+\CO{J_0}{\AC{G_s}{G'_r}}\nonumber\\
& = -\AC{G_s}{G_r}-(s-r)\CO{J^R_0}{J^R_{s+r}} \nonumber\\
& =  -\AC{G_s}{G_r}~.
\end{align}
These commutation relations, together with~(\ref{eq:N1modes}) translate to
\begin{align}
\label{eq:necessaryN2}
\CO{J^R_m}{J^R_n} &= \ff{c}{3} m\delta_{m,-n}~,  \quad \CO{J^R_n}{G^\pm_r} = \pm G^\pm_{n+r}~, \quad \AC{G^\pm_r}{G^\pm_s} = 0~, \nonumber\\[5mm]
\AC{G^+_r}{G^-_s} & = 2 L_{r+s} +(r-s) J_{r+s} + \ff{c}{3}(r^2-\ff{1}{4}) \delta_{r,-s}~
\end{align}
with $c=12$.  Jacobi identities then lead to the second set of commutators
\begin{align}
\label{eq:remainingN2}
\CO{L_m}{G^\pm_r} & = (\ff{m}{2}-r)G^\pm_{m+r}~, \qquad
\CO{L_m}{J^R_n} = -n J^R_{m+n}~, \nonumber\\[5mm]
\CO{L_m}{L_n} & = (m-n)L_{m+n} + \ff{c}{12} (m^3-m)\delta_{m,-n}~.
\end{align}
The two sets of relations~(\ref{eq:necessaryN2}) and~(\ref{eq:remainingN2}) are the defining relations of the $\cA^2_{12}$ SCA.

So, we have shown that $\cN =(2,0)$ spacetime supersymmetry requires the internal theory to have a $\cA^2_{12}$ SCA.  Moreover, the spin fields are constructed by bosonizing the R-current and carry charges $\pm 1$.  It follows that locality of the OPE requires all NS sector operators to carry integral R-charge, with odd charges charges for fermions and even ones for bosons.

\subsection{$\cN=(3,0)$}
With $p>2$ spin fields $\Sigma^A_+$ the worldsheet theory has an $\widehat{\so(p)}_1$ Kac-Moody algebra in the NS sector, with currents $J^{AB} = i \Sigma^A \Sigma^B$ for $A>B$.  We can decompose these in representations of the $\cA^2_{12}$ SCA, and in particular with respect to the $\Lu(1)_R$ Kac-Moody:
\begin{align}
\so(p) & \supset \so(p-2) \oplus \so(2)_{R}~, \nonumber\\
\wedge^2 \rep{p} & = \wedge^2(\rep{p-2})_{0}  \oplus \rep{(p-2)}_{+2} \oplus \rep{(p-2)}_{-2} \oplus \rep{1}_{0}~.
\end{align}
We let $\alpha = 3,\ldots, p$, and denote the currents in the decomposition as 
$J^{\alpha\beta}$, $\cJ^\alpha$,  $\cJ^{\dag \alpha}$, and $J^R$.
The non-zero OPEs for these are
\begin{align}
\label{eq:bigKM}
\qquad
J^R_1 J^R_2  &\sim \frac{4}{z_{12}^2}~,\nonumber\\
J^{\alpha\beta}_1 J^{\gamma\delta}_2 & 
\sim \frac{1}{z_{12}^2}  \left[\delta^{\alpha\gamma} \delta^{\beta\delta}-\delta^{\alpha\delta} \delta^{\beta\gamma}\right]
+\frac{i}{z_{12}} \left[\delta^{\alpha\delta} J^{\beta\gamma} - \delta^{\alpha\gamma} J^{\beta\delta}_2 + \delta^{\beta\gamma} J^{\alpha\delta}_2 -\delta^{\beta\delta} J^{\alpha\gamma}_2\right]~,
\nonumber\\
J^{\alpha\beta}_1 \cJ^{\gamma}_2 & \sim \frac{i}{z_{12}}\left[  \delta^{\beta\gamma} \cJ^\alpha_2-\delta^{\alpha\gamma} \cJ^\beta_2 \right]~,
\qquad
J^{\alpha\beta}_1 \cJ^{\dag\gamma}_2  \sim \frac{i}{z_{12}}\left[ \delta^{\beta\gamma} \cJ^{\dag\alpha}_2 - \delta^{\alpha\gamma} \cJ^{\dag\beta}_2\right]~,
\nonumber\\
\cJ^{\alpha}_1\cJ^{\dag \beta}_2 & \sim \frac{2\delta^{\alpha\beta}}{z_{12}^2} +  \frac{1}{z_{12}} \left[-2i J^{\alpha\beta}+ \delta^{\alpha\beta} J^R_2 \right]~,\quad
J^R_1 \cJ^{\alpha}_2  \sim \frac{+2 \cJ^\alpha_2}{z_{12}}~,\quad
J^R_1 \cJ^{\dag\alpha}_2  \sim \frac{-2 \cJ^{\dag\alpha}_2}{z_{12}}~.
\end{align}
From the last line it follows that the currents $\cJ^{\alpha}$ are chiral primary with respect to $\cA^2_{12}$, while the $\cJ^{\dag\alpha}$ are their anti-chiral primary conjugates.  Moreover, for any $\alpha$
the triplet $\{\cJ^\alpha,\cJ^{\dag\alpha},J^R\}$ forms an $\widehat{\su(2)}_2$ Kac-Moody algebra.
As a result, we have the additional non-trivial OPEs
\begin{align}
G^+_1 \cJ^\alpha_2 & \sim O(1)~,&
G^-_1 \cJ^\alpha_2 & \sim -\frac{ \cG^\alpha}{z_{12}}~,&
G^+_1 \cG^\alpha_2 & \sim -\frac{4 \cJ^\alpha_2}{z_{12}^2} - \frac{2 \p \cJ^\alpha_2}{z_{12}}~,\nonumber\\
G^-_1 \cJ^{\dag\alpha}_2 &\sim O(1)~, &G^+_1 \cJ^{\dag \alpha}_2 & \sim \frac{ \cG^{\dag\alpha}}{z_{12}}~, &
G^-_1 \cG^{\dag\alpha}_2 & \sim  \frac{4 \cJ^{\dag\alpha}_2}{z_{12}^2} + \frac{2 \p \cJ^{\dag\alpha}_2}{z_{12}}~.
\end{align}
where $\cG^\alpha$ is primary with respect to Virasoro and $\widehat{\Lu(1)}_R$ and has $h=3/2$ and $q=+1$, and
$\cG^{\dag\alpha}$ is conjugate to $\cG^\alpha$.  

If $\cN = (3,0)$, then $J^{\alpha\beta} = 0$, and the $N=2$ $c=12$ theory contains a chiral primary multiplet with $q=2$ and fields $\cJ,\cG$, as well as the conjugate anti-chiral primary multiplet.  A straightforward application of Jacobi identities shows that $G^\pm$ and $\cG,\cG^\dag$ arrange themselves into doublets of the $\widehat{\su(2)}_2$, and the $N=2$ SCA is enlarged to the small $N=4$ SCA, which we denote by $\cA^4_{12}$.    Thus, we have shown that for $\cN\ge(3,0)$ we have $\cA^2_{12} \subset \cA^4_{12}$.

A geometric example of this is provided by compactification on a manifold with $\Sp(2)$ holonomy, where the triplet of hyper-K\"ahler forms leads to the $\widehat{\su(2)}_2$ algebra.

\subsection{$\cN=(4,0)$}
We will show that if $\cN \ge (4,0)$ then $\cA^4_{12} \subset \cA^4_6\oplus \cA^4_6$.  To prove this, we at first forget about the full $N=4$ algebra and just consider the $N=2$ structure.  When $p \ge 4$, in addition to $J^R = 2i \Sigma^1 \Sigma^2$, we have another commuting $\Lu(1)$ current $J^{34} = i \Sigma^3\Sigma^4$, and we decompose the supercharges $G^\pm$ further with respect to $J^{34}$.  As above, locality of the OPE implies that the only allowed charges in the decomposition are $q_{34} = \pm 1/2$.  It follows that if we write
\begin{align}
\bJ^1 &= \ff{1}{2}J^R + J^{34}~, & 
\bJ^2 & = \ff{1}{2}J^R - J^{34}~
\end{align}
and decompose the supercharges with respect to $\bJ^1,\bJ^2$ charges, we must have
\begin{align}
G^+ & = \bG^{+1,0} + \bG^{0,+1}~, &
G^- & =  \bG^{-1,0} + \bG^{0,-1}~.
\end{align}
As we will soon see, the $\bG^{\pm 1,0}$ and $\bG^{0,\pm 1}$ lead to two commuting $N=2$ SCAs, each with $c=6$.

\subsubsection*{The long $N=2$ multiplet}
The operators $\bJ^1$ and $\bG^{\pm 1,0}$ are completed to a long $N=2$ quasi-primary multiplet by an additional spin $2$ operator $\bT^1$, and the OPEs are 
\begin{align}
\label{eq:longN2OPE}
\bJ^1_1 G^{\pm}_2 & \sim \pm \frac{\bG^{\pm 1,0}}{z_{12}}~,  \nonumber\\
G^{\pm}_1 \bG^{\pm 1,0}_2 &\sim O(1)~, \nonumber\\
G^+_1 \bG^{-1,0}_2 & \sim \frac{2 c/3}{z_{12}^3} + \frac{2 \bJ^1_2}{z_{12}^2} + \frac{2 \bT^1_2 + \p \bJ^1_2}{z_{12}}~, \nonumber\\
G^-_1 \bG^{+1,0}_2 & \sim \frac{2c/3}{z_{12}^3}-\frac{2 \bJ^1_2}{z_{12}^2} + \frac{2 \bT^1_2 - \p \bJ^1_2}{z_{12}}~, \nonumber\\
\bT^1_2 G^{\pm}_2 & \sim \frac{3/2 \bG^{\pm 1,0}_2}{z_{12}^2} + \frac{\p \bG^{\pm 1,0}_2}{z_{12}}~, \nonumber\\
\bT^1_2 J^R_2 & \sim \frac{\bJ^1_2}{z_{12}^2} + \frac{ \p\bJ^1_2}{z_{12}}~,\nonumber\\
T_1 \bT^1_2 & \sim \frac{c/2}{z_{12}^4} + \frac{ 2\bT^1_2}{z_{12}^2} + \frac{\p\bT^1_2}{z_{12}}~,
\end{align}
where $c=6$.  There are analogous relations for  $\bJ^2$ , $\bG^{0,\pm 1}$ and a spin 2 operator $\bT^2$.
Consider the following expressions implied by these OPEs:
\begin{align}
\bG^{\pm1,0}_1 \bG^{\pm 1,0}_2 + \bG^{0,\pm 1}_1\bG^{\pm1,0}_2  &\sim O(1)~, \nonumber \\
\bG^{1,0}_1 \bG^{-1,0}_2 + \bG^{0,1}_1\bG^{-1,0}_2 - \bG^{-1,0}_1 \bG^{+1,0}_2 - \bG^{0,-1}_1 \bG^{+1,0}_2
&\sim \frac{4 \bJ^1_2}{z_{12}^2} +\frac{2\p\bJ^1_2}{z_{12}}~. 
\end{align}
We can grade these by the global charges corresponding to $\bJ^1$ and $\bJ^2$, and since the right-hand-sides are neutral under both of these, we conclude that
\begin{align}
\bG^{\pm 1,0}_1 \bG^{\pm 1,0}_2 &\sim O(1)~,&
\bG^{0,\pm 1}_1 \bG^{\pm 1,0}_2 &\sim O(1)~,&
\bG^{0,\pm 1}_1 \bG^{\mp 1,0}_2 &\sim O(1)~.
\end{align}
So, the $\bG^{\pm1,0}$ supercharges anticommute with the $\bG^{0,\pm 1}$ supercharges.  In that case we
see from~(\ref{eq:longN2OPE}), ~(\ref{eq:necessaryN2}) and ~(\ref{eq:remainingN2})
 that $\bJ^1,\bG^{\pm 1,0}, \bT^1$ and $\bJ^2,\bG^{\pm 2,0}, \bT^2$ define two commuting $\cA^2_{6}$ algebras, and their sum yields the $\cA^2_{12}$---in particular, $T = \bT^1 + \bT^2$.

We have now shown that if $\cN \ge (4,0)$ then $\cA^2_{12} \subset \cA^2_6 \oplus \cA^2_6$.  In addition, the currents of  the $\widehat{\so(4)}_1 = \widehat{\su(2)}_1 \oplus \widehat{\su(2)}_1$ KM algebra decompose with respect to the N=2 R-symmetry $\widehat{\Lu(1)}_2\oplus\widehat{\Lu(1)}_2$ as
\begin{align}
\rep{6} = \rep{1}_{+2,0} \oplus \rep{1}_{0,0} \oplus \rep{1}_{-2,0} \oplus \rep{1}_{0,+2} \oplus \rep{1}_{0,0} \oplus \rep{1}_{0,-2}~.
\end{align}
Thus, each of the $\cA^{2}_6$ factors includes a chiral primary field with $\bq = 2$:  the corresponding short multiplets and their anti-chiral conjugates contain the additional supercharges that enhance each of the $\cA^2_6$ factors to $\cA^{4}_6$.  Since the spectrum is then organized into $\widehat{\su(2)}_1 \oplus \widehat{\su(2)}_1$ representations, unitarity implies that R-charge integrality holds in the NS sector for each of the $\cA^4_6$ factors separately, and that means that a worldsheet theory with $\cA^{4}_6\oplus \cA^{4}_6$ leads to $\cN \ge (4,0)$ spacetime supersymmetry.  A geometric realization is provided by compactification on $\text{K3}\times\text{K3}$.

\subsection{Beyond $\cN = (4,0)$} \label{ss:bigsusy}
For $p>4$ the current algebra~(\ref{eq:bigKM}) contains $\widehat{\so(5)}_1$ as a subalgebra.  We can decompose it with respect to the structure already obtained for $p=4$ as
\begin{align}
\widehat{\so(5)}_1 \supset \widehat{\su(2)}_1 \oplus \widehat{\su(2)}_1 \supset \widehat{\Lu(1)}_2 \oplus \widehat{\Lu(1)}_2~,
\end{align}
where the last two factors are generated by $\bJ^1$ and $\bJ^2$.  The $\so(5)$ currents decompose as
\begin{align}
\rep{10} = (\rep{3},\rep{1})\oplus(\rep{1},\rep{3})\oplus(\rep{2},\rep{2})~,
\end{align}
and now for the interesting part: under the R-symmetries we have the decomposition
\begin{align}
\label{eq:so5currents}
(\rep{2},\rep{2}) = \rep{1}_{+1,+1} \oplus \rep{1}_{+1,-1} \oplus \rep{1}_{-1,+1} \oplus \rep{1}_{-1,-1}~.
\end{align}
The first term has $h=1$ and $q^R = \bq_1 + \bq_2  = 2$ (the last term is then its conjugate).  In other words, the $\rep{1}_{+1,+1}$ current is chiral primary with respect to the diagonal $N=2$ algebra, while the $\rep{1}_{-1,-1}$ current is anti-chrial primary.  Since our algebra decomposes into two $N=2$ algebras, this means that these conditions hold separately with respect to each of the algebras.\footnote{An operator is chiral primary if and only if it is annihilated by $G^+_{-1/2}$ and $G^{\pm}_r$ operators with $r>0$.  Since the supercharges decompose, an operator that is chiral primary with respect to the diagonal algebra must be chiral primary with respect to each subalgebra. }  So, the operators are represented as
\begin{align}
\rep{1}_{+1,+1}  & \leftrightarrow \bpsi^1 \bpsi^2~, &
\rep{1}_{-1,-1}  & \leftrightarrow \bpsi^{1\dag} \bpsi^{2\dag}~,
\end{align}
where $\bpsi^1$ is chiral primary with $\bq_1 = 1$, $\bq_2 = 0$, while $\bpsi^{1\dag}$ is anti-chiral primary with $\bq_1 = -1$ and $\bq_2 = 0$, and similarly for $\bpsi^{2}$ and $\bpsi^{2\dag}$.  In order to have the correct $\widehat{\so(5)}_1$ commutation relations the operators need to satisfy
\begin{align}
\bpsi^a(z_1) \bpsi^b(z_2) &= O(z_{12})~,&
\bpsi^a(z_1) \bpsi^{b\dag}(z_2) \sim \frac{\delta^{ab}}{z_{12}}~.
\end{align}
for $a,b\in\{1,2\}$.  

Suppose for a moment that the operators $\bpsi^{1}$ and $\bpsi^{2}$ are present in the SCFT spectrum and focus on the first $c=6$ SCA (analogous statements will hold for the second one).  From what we just said, it has a chiral multiplet with components $\bpsi^1$ and its superpartner $j^{1}$, a holomorphic current; there is also an anti-chiral conjugate multiplet with components $\bpsi^{1\dag }$ and $j^{1\dag}$.  The singular OPEs are
\begin{align}
\bpsi^1_1 \bpsi^{1\dag}_2 &\sim \frac{1}{z_{12}}~, &
j^{1\dag}_1 j^1_2 & \sim \frac{1}{z_{12}^2}~.
\end{align}
It is then straightforward to show that (we are leaving out normal ordering)
\begin{align}
\label{eq:freecthree}
\bj^1 & = \bpsi^1 \bpsi^{1\dag}~, &
\bg^{1+}& = \sqrt{2} \bpsi^1 j^{\dag1}~,&
\bg^{1-} &= \sqrt{2}\bpsi^{1\dag} j^{1}~,&
\bt^{1} & = j^1 j^{1\dag} -\ff{1}{2} (\bpsi^{1}\p \bpsi^{1\dag} + \bpsi^{1\dag}\p \bpsi^{1}) 
\end{align}
generate a $c=3$ $N=2$ SCA, and the $c=6$ $N=2$ SCA decomposes into a sum of two commuting
$c=3$ SCAs---this is just the N=2 super-Sugawara construction. Moreover, we also know from the preceding section that $\bq_1,\bq_2 \in \Z$ for all NS sector states.  This means that R-charge integrality holds for each of the $\cA^2_3$ factors in $\cA^2_6 \subset \cA^2_3\oplus\cA^2_3$, so that each factor has the representation of the form~(\ref{eq:freecthree}).  Exactly the same thing happens to the second $c=6$ $N=2$ SCA, so that if the fermions $\bpsi^a$ are in the spectrum, then the fermionic content of the holomorphic NS sector consists of $4$ Weyl fermions.  But, we know that in this case we obtain $16$ spin fields---$8$ with $e^{i\pi F} = 1$, and $8$ with $e^{i\pi F} = -1$; these lead to $\cN = (8,8)$ spacetime supersymmetry.

However, there is also the possibility that the operators $\bpsi^{a}$ are not in the CFT spectrum---for instance they may have been projected out by an orbifold action that nevertheless preserves the currents represented by $\bpsi^1\bpsi^2$.\footnote{This is the point we missed in the earlier version of the paper that was made clear by the orbifold constructions in~\cite{Florakis:2017zep}.}   So, in general, we can only conclude that a chiral $\cN = (p,0)$ spacetime supersymmetry with $p\ge 4$ requires a worldsheet SCA $\cA^{4}_6\oplus \cA^{4}_6$, where the R-symmetry is a subalgebra of a larger Kac-Moody symmetry:
\begin{align}
\widehat{\so(p)}_1 &\supset \widehat{\su(2)}_1 \oplus \widehat{\su(2)}_1  \oplus \widehat{\so(p-4)}_1 \nonumber\\
\rep{p} & = (\rep{2},\rep{2},\rep{1}) \oplus (\rep{1},\rep{p-4})~,\nonumber\\
\wedge^2\rep{p} &= (\rep{3},\rep{1},\rep{1})\oplus(\rep{1},\rep{3},\rep{1})\oplus(\rep{2},\rep{2},\rep{p-4})~.
\end{align}
The currents in the $(\rep{2},\rep{2},\rep{p-4})$ representation form short $h=1/2$ multiplets with respect to each of the $\cA^4_6$ factors.  While we clearly must have $p \le 24$ to be consistent with $c=12$, it is not easy to see any further restrictions on $p$ that follow from representation theory.  Further conditions may follow from modular invariance.

\section{(p,q) spacetime supersymmetry and lifts to three dimensions} \label{s:pq}
We now generalize to $\cN = (p,q)$ with both $p$ and $q$ non-zero.  Without loss of generality we assume $p\ge q \ge 1$.  It turns out that the analysis is straightforward for all cases except the ``basic one'' of $\cN = (1,1)$, which we defer to the following section.  

\subsection{$\cN = (2,1)$}
From the previous section we know that the internal theory has a $\cA^2_{12}$ SCA.  We also know from section~\ref{ss:ramond} that the internal theory has at least one NS sector fermion $\Psi$ with $h=1/2$.  In a unitary compact theory such a fermion must be an $N=2$ primary field.\footnote{This is so because by weight it can only be a descendant of the vacuum, but that is not possible in a unitary compact theory.}  Every operator in the NS sector satisfies the unitarity bound $h\ge |q|/2$, where $q$ is the R-charge.  But, we also have the constraints from locality:   the $\cA^2_{12}$ NS representations have $q\in \Z$, and the charge is correlated with fermion number.  Therefore, $\Psi$ must be a linear combination of a chiral primary field $\bpsi$ with $q=1$ and an anti-chiral primary field with $q= -1$.

This is sufficient to conclude that $\cA^2_{12} \subset \cA^2_{9} \oplus \cA^2_{3}$, as might be expected from the geometric case $X_8 = X_6 \times T^2$ described above.  The basic reason is that $\bpsi$ forms a short $N=2$ multiplet, and the super-Kac-Moody construction illustrated in~(\ref{eq:freecthree}) implies that we can decompose the SCA into two commuting factors, with the second associated to the free fermion and its partner bosonic currents.  Since R-charge integrality holds for the $\cA^2_{12}$ factor and, obviously, for the free fermion $\cA^2_3$ factor, it must also hold for the $\cA^2_9$ commutant.  This means we have spectral flow in each of the factors separately, and, just as in the non-chiral example in section~\ref{ss:geometryexamples}, we obtain $\cN = (2,2)$.   The bottom line is that $\cN=(2,1)$ implies $\cN= (2,2)$.

\subsection{$\cN = (3,2)$}
With $\cN=(3,2)$ supersymmetry we know that, on the one hand $\cA^{2}_{12} \subset \cA^{2}_9\oplus \cA^{2}_3$, with R-current decomposing as $J^R = \bJ + \bpsi \bpsi^\dag$, while on the other hand $\cA^{2}_{12} \subset \cA^{4}_{12}$.  The first statement means there is a holomorphic fermion $\bpsi$ and its superpartner $j$, and these, together with their conjugates generate the $\cA^{2}_3$ SCA as in~(\ref{eq:freecthree}).  The second statement implies that $\bpsi$ is in a non-trivial representation of $\widehat{\su(2)}_2$, and, moreover, the charge $+2$ current $\cJ$ of the $\widehat{\su(2)}_2$ is chiral primary with respect to $\cA^{2}_9$ and $\cA^{2}_3$ SCAs, i.e. we can write
\begin{align}
\cJ = \cK + \chi \bpsi~,
\end{align}
where $\cK$ and $\chi$ are chiral primary fields with respect to $\cA^{2}_9$, with R-charges $+2$ and $+1$ respectively, and similarly for the conjugate anti-chiral primary current  $\cJ^\dag$.  The OPE
\begin{align}
\cJ_1 \cJ^\dag_2 &\sim \frac{2}{z_{12}^2}  + \frac{1}{z_{12}} \left( \bJ + \bpsi\bpsi^\dag\right)_2
\end{align}
implies that the $\chi\bpsi$ term in $\cJ$ must be non-zero:  otherwise it is not possible to obtain the $\bpsi\bpsi^\dag$ term from the OPE.  Moreover, since $\bpsi$ is in a non-trivial representation of $\widehat{\su(2)}_2$, $\chi$ is indeed present in the spectrum, since
\begin{align}
\cJ_1 \bpsi^\dag_2 \sim \frac{1}{z_{12}} \chi_2~.
\end{align}
But now we can apply exactly the same arguments as in the previous section to conclude that $\cA^2_9 \subset \cA^2_6\oplus \cA^2_3$, where the $\cA^2_3$ factor is associated to the free fermions $\chi$, $\chi^{\dag}$ and their superpartners.  The current $\cK$ is then chiral primary with respect to $\cA^2_6$, and it, together with $\cK^\dag$ complete the algebra to $N=4$. 

It follows that $\cN(3,2)$ spacetime supersymmetry requires  $\cA^2_{12} \subset \cA^{4}_6\oplus\cA^2_3\oplus\cA^2_3$, where the last two factors are associated to the free Weyl fermions $\chi$ and $\bpsi$.  This structure and R-charge integrality imply $\cN=(4,4)$ spacetime supersymmetry, and it has the geometric realization as a $\text{K3}\times T^4$ compactification.

\subsection{$\cN\ge (5,4)$}
We now reconsider the $\widehat{\so(5)}_1$ currents of~(\ref{eq:so5currents}) in light of the SCA decomposition $\cA^{4}_6\oplus \cA^2_3\oplus \cA^2_3$. In this case $\bpsi^2$ of~(\ref{eq:so5currents}) is a linear combination of the free fermions in the $\cA^{2}_3\oplus\cA^2_3$ factor.  That in turn means that $\bpsi^1$ is in the spectrum and is chiral primary with respect to $\cA^4_6$ with $\bq_1=1$.   It follows that $\cA^4_6$ also decomposes as $\cA^2_3\oplus\cA^2_3$, where R-charge integrality holds for each factor.\footnote{As above, this follows because $\cA^4_6$ has integral R-charges by unitarity of the $\widehat{\su(2)}_1$ KM algebra, and one of the $\cA^2_3$ factors is generated by the free fermion $\bpsi^1$ and its superpartners and conjugates and obviously has integral R-charges in the NS sector, while the other  $\cA^2_3$ factor is its commutant, and the full spectrum must be integral with respect to its R-symmetry.}  So, all in all, the full $c=12$ SCA now decomposes as $(\cA^2_3)^{\oplus 4}$, the holomorphic SCA of the $T^8$ conformal theory.  It follows that $\cN \ge (5,4)$ implies $\cN = (8,8)$.

\section{The tri-critical Ising model and $\cN= (1,1)$ supersymmetry}
We now come to what is probably the most intricate worldsheet structure in this classification.  It was uncovered in the weakly coupled non-linear sigma model with a $\GG_2$ holonomy target space in~\cite{Shatashvili:1994zw}. 
We will show that this structure must be present to realize $\cN=(1,1)$ supersymmetry in $D=2$.

From sections~\ref{ss:ramond} and~\ref{ss:minsusy} we know that the worldsheet theory will have have N=1 superconformal invariance, and the NS sector has two interesting features: there is a Majorana-Weyl fermion $\Psi$ with spin $1/2$, and there is a spin $2$ operator $T^{\Sigma}= -\ff{1}{2} \Sigma_+ \p \Sigma_+$, which is the energy momentum tensor of the Ising model.  Let $T$ be the full energy momentum tensor of the theory.  We then have
\begin{align}
\label{eq:TSigma}
T_1 T^{\Sigma}_2 & \sim \frac{1/4}{z_{12}^4} + \frac{ 2T^\Sigma_2}{z_{12}^2} + \frac{\p T^\Sigma_2}{z_{12}}~, \nonumber\\
T^\Sigma_1 T^{\Sigma}_2 & \sim \frac{1/4}{z_{12}^4} + \frac{ 2T^\Sigma_2}{z_{12}^2} + \frac{\p T^\Sigma_2}{z_{12}}~, \nonumber\\
\end{align}
as well as
\begin{align}
\label{eq:PsiPsi}
\Psi_1 \Psi_2 & =  \frac{1}{z_{12}} + 2 z_{12} T_2^\Psi + z_{12}^2 \p T_2^\Psi+O(z_{12}^3)~,
\end{align}
where $T^\Psi = -\ff{1}{2} \Psi \p \Psi$ is the energy momentum tensor for another Ising model.

We decompose the full energy-momentum tensor $T$ into two commuting summands, $T = T' + T^{\Psi}$, where $T'$ generates a Virasoro algebra with $c=23/2$.  With respect to this decomposition the operator $T^{\Sigma}$ has the form
\begin{align}
T^{\Sigma} = X+ \Psi \Phi + \p \Psi \Xi + \xi T^{\Psi}~,
\end{align}
where $X$, $\Phi$, $\Xi$ are quasi-primary operators of weight, respectively, $h' = 2$, $h'=3/2$, and $h'=1/2$, while $\xi$ is a constant.  If the internal theory contains an additional free fermion $\Xi$ in the NS sector, then the resulting theory will have enhanced spacetime supersymmetry compared to the case we are after; in fact it must have at least $\cN = (2,2)$.  Therefore, we will restrict attention to the minimal situation, where $\Xi = 0$ and 
\begin{align}
\label{eq:TSigmaAnsatz}
T^{\Sigma} = X+ \Psi \Phi + \xi T^{\Psi}~,
\end{align}
The OPEs of $X$ and $\Phi$ are constrained by~(\ref{eq:TSigma}), and we will show that they lead to the appearance of the tricritical Ising model, with $\Phi$ proportional to its  supercharge and $X$ proportional to its energy-momentum tensor.

\subsection*{Determination of the constant $\xi$}
As a first step, we find the value of the constant $\xi$.  To do this, we decompose the spin field $\Sigma_+$ with respect to the $T'+T^{\Psi}$ decomposition as
\begin{align}
\Sigma_+ = \Sigma' \sigma~,
\end{align}
where $\sigma$ is the Ising spin field with $h=1/16$, and $\Sigma'$ has $h' = 7/16$.\footnote{Note that $\sigma$ is the spin field with respect to the the $\Psi$ Ising model, not the $\Sigma_+$ one!}  That, together with conformal invariance and $\la \Sigma_{+1} \Sigma_{+2}\ra = 1/z_{12}$ fixes the three-point function
\begin{align}
\la T^\Psi \Sigma_{+2} \Sigma_{+3} \ra = \frac{1}{16} \frac{z_{23}}{z_{12}^2 z_{13}^2}~.
\end{align}
But then we have, on one hand, from~(\ref{eq:TSigmaAnsatz})
\begin{align}
\la T^{\Psi}_1 T^\Sigma_3 \ra = \frac{\xi}{4} \frac{1}{z_{13}^4}~,
\end{align}
while on the other hand,
\begin{align}
\la T^{\Psi}_1 T^\Sigma_3 \ra = - \frac{1}{2} \lim_{z_2 \to z_3}  \p_3 \la T^\Psi \Sigma_{+2} \Sigma_{+3} \ra = \frac{1}{32} \frac{1}{z_{13}^4}~.
\end{align}
So, we conclude that $\xi = 1/8$.

\subsection*{Constraints on $\Phi$ and $X$}
We obtain further constraints by plugging~(\ref{eq:TSigmaAnsatz}) with $\xi = 1/8$ into~(\ref{eq:TSigma}).  From the $T_1 T^\Sigma_2$ OPE we conclude
\begin{align}
T'_1 \Phi_2 &\sim \frac{3/2 \Phi_2}{z_{12}^2} + \frac{\p \Phi_2}{z_{12}}~, &
T'_1 X_2 & \sim \frac{7/32}{z_{12}^4} + \frac{2 X_2}{z_{12}^2} +\frac{\p X_2}{z_{12}}~.
\end{align}
Furthermore, the most general OPE structure for $\Phi$ and $X$ consistent with our assumptions
on the internal theory is
\begin{align}
\Phi_1\Phi_2&\sim \frac{a}{z_{12}^3} + \frac{2\Xt_2}{z_{12}} +\p \Xt_2~, \nonumber\\[1mm]
X_1 \Phi_2 & \sim  \frac{\Phit_2}{z_{12}^2} + \frac{\Pi_2}{z_{12}}~, \nonumber\\
X_1 X_2 & \sim \frac{b}{z_{12}^4} + \frac{2 Y_2}{z_{12}^2} +\frac{\p Y_2}{z_{12}}~.
\end{align}
Here $a$ and $b$ are undetermined constants, while $\Xt$ and $Y$ are operators with $h' = 2$, and $\Phit$ and $\Pi$ have, respectively, $h' = 3/2$ and $h' = 5/2$.\footnote{The assumptions are the spin/statistics relation, and the absence of NS fields with $h' = 1/2$.}  Using this in the $T^\Sigma_1 T^\Sigma_2$ OPE constrains the unknown
operators and constants; we find $a =-7/64$, $b= 35/256$, $\Phit = 15/16 \Phi$, and $Y = X +\Xt$, so that our Ansatz takes the form
\begin{align}
\label{eq:OPEalmost}
T^{\Sigma} & = \Psi \Phi+ \ff{1}{8} T^{\Psi} + X~, \nonumber\\
\Phi_1\Phi_2&\sim \frac{-7/64}{z_{12}^3} + \frac{2\Xt_2}{z_{12}} +\p \Xt_2~, \nonumber\\[1mm]
X_1 \Phi_2 & \sim  \frac{15/16\Phi_2}{z_{12}^2} + \frac{\Pi_2}{z_{12}}~, \nonumber\\
X_1 X_2 & \sim \frac{35/256}{z_{12}^4} + \frac{2 (X_2+\Xt_2)}{z_{12}^2} +\frac{\p (X_2+\Xt_2)}{z_{12}}~.
\end{align}
It would non-trivial to proceed without the insights of~\cite{Shatashvili:1994zw}; but since we know we are aiming to find a tri-critical Ising structure, we see how to organize these results.  Define
\begin{align}
\Gt & = \frac{8i}{\sqrt{15}} \Phi~, &
\Tt & = \frac{8}{5} X~, &
Z   & = \Xt + \frac{3}{8} X~,&
\cX & = \Pi - \frac{5}{8} \p \Phi~.
\end{align}
Now~(\ref{eq:OPEalmost}) is rewritten as
\begin{align}
T^{\Sigma} & = \frac{\sqrt{15}}{8i} \Psi \Gt + \ff{1}{8} T^{\Psi} +\frac{5}{8} \Tt~,\nonumber\\
\Gt_1 \Gt_2 &\sim \frac{2C/3}{z_{12}^3} + \frac{ 2 \Tt_2}{z_{12}}~ + \frac{Z_2}{z_{12}},  \nonumber\\
\Tt_1 \Gt_2 & \sim \frac{ 3/2 \Gt_2}{z_{12}^2} + \frac{ \p\Gt_2}{z_{12}} +\frac{ \cX_2}{z_{12}}~,\nonumber\\
\Tt_1 \Tt_2 & \sim \frac{ C/2}{z_{12}^4} + \frac{2 \Tt_2}{z_{12}^2} + \frac{\p \Tt_2}{z_{12}} 
-\frac{3}{5}\left( \frac{2Z_2}{z_{12}^2}
+\frac{\p Z_2}{z_{12}}\right)~,\nonumber\\
T'_1 \Tt_2 & \sim \frac{ C/2}{z_{12}^4} + \frac{2 \Tt_2}{z_{12}^2} + \frac{\p \Tt_2}{z_{12}}~,\nonumber\\
T'_1 \Gt_2 & \sim \frac{3/2 \Gt_2}{z_{12}^2} + \frac{\p \Gt_2}{z_{12}}~,
\end{align}
where $C = 7/10$.  If $\cX$ and $Z$ vanish, then the Virasoro algebra decomposes into two commuting summands, one of which is the tri-critical Ising model with currents $\Tt$ and $\Gt$.

\subsection*{The vanishing of the ``remainders'' $\cX$ and $Z$}
The operators $\cX$ and $Z$ do vanish.  To demonstrate this, we consider the mode expansion
\begin{align}
\AC{\Gt_r}{\Gt_s} & = \ff{C}{3}(r^2-\ff{1}{4})\delta_{r,-s} + 2 \Lt_{r+s} + 2 Z_{r+s}~,\nonumber\\
\CO{\Lt_m}{\Gt_s} & = (\ff{m}{2}-s) \Gt_{m+s} + \cX_{m+s}~, \nonumber\\
\CO{\Lt_m}{\Lt_n} & = \ff{C}{12} (m^3-m) \delta_{m,-n} + (m-n) \Lt_{m+n} -\ff{3}{5}(m-n)Z_{m+n}~,\nonumber\\
\CO{L'_m}{\Lt_n} & = \ff{C}{12} (m^3-m) \delta_{m,-n} + (m-n) \Lt_{m+n} ~,\nonumber\\
\CO{L'_m}{\Gt_s} & = (\ff{m}{2}-s) \Gt_{m+s}~.
\end{align}
The Jacobi identities lead to a number of constraints which include
\begin{align}
\label{eq:bigJacobi}
\Gt\Gt\Gt &:& &3\cX_{r+s+t}  = \CO{\Gt_t}{Z_{r+s}}+\CO{\Gt_s}{Z_{t+r}}+\CO{\Gt_r}{Z_{s+t}}~, \nonumber\\[2mm]
\Lt\Gt\Gt &:&  &\CO{\Lt_p}{Z_{r+s}} = \ff{1}{2} \left\{\AC{\Gt_r}{\cX_{p+s}} + \AC{\Gt_s}{\cX_{p+r}}\right\} + \ff{8}{5} Z_{p+r+s}~,\nonumber\\[2mm]
L' \Gt\Gt  &:& & \CO{L'_m}{Z_n}  = (m-n) Z_{m+n}~.
\end{align}
Setting $r=s=t=-1/2$ and $p=-1$, we obtain
\begin{align}
\label{eq:theend}
\cX_{-3/2} & = \CO{\Gt_{-1/2}}{Z_{-1}}~, \nonumber\\
\CO{\Lt_{-1}}{Z_{-1}} & = \AC{\Gt_{-1/2}}{\cX_{-3/2}} + \ff{8}{5} Z_{-2}~.
\end{align}
But, starting with the first line, we also have
\begin{align}
\AC{\Gt_{-1/2}}{\cX_{-3/2}} = \AC{\Gt_{-1/2}}{\CO{\Gt_{-1/2}}{Z_{-1}}} = \ff{1}{2} \CO{\AC{\Gt_{-1/2}}{\Gt_{-1/2}}}{Z_{-1}} = \CO{\Lt_{-1}}{Z_{-1}}~.
\end{align}
The second equality follows from the Jacobi identity, and the third one from the anti-commutator above.  Using that in the second line of~(\ref{eq:theend}), we obtain
\begin{align}
Z_{-2} = 0~.
\end{align}
Since $Z$ is a Virasoro primary operator with $h' = 2$, we now see that $Z=0$ as an operator, since the state $\lim_{z\to 0} Z(z) |0\ra = Z_{-2} |0\ra = 0$.  Once this holds, then the first line of~(\ref{eq:bigJacobi}) implies $\cX =0$ as well.

\bibliographystyle{./utphys}
\bibliography{./bigref}

\end{document}